# Similarity of structuring in the range $10^{-5}$ cm to $10^{23}$ cm hints at a baryonic cold dark skeleton of the Universe

A.B. Kukushkin & V. A. Rantsev-Kartinov

*RRC "Kurchatov Institute", Moscow, 123182, Russia*

The presence of skeletal structures of the same distinctive topology (cartwheels, tubules, etc.), in the range $10^{-5}$-$10^{23}$ cm, and a trend toward self-similarity of these structures are found. These evidences come from the electron micrography of dust deposits in tokamak ($10^{-6}$-$10^{-3}$ cm), the high resolution images of plasma taken in the visible light and soft x-rays in laboratory electric discharges -- tokamaks, Z-pinches, plasma focus and vacuum spark ($10^{-2}$-$10$ cm), hail particles (1-10 cm), the images of severe weather phenomena -- tornado ($10^{3}$-$10^{5}$ cm), modern databases (the Hubble Space Telescope's and Chandra X-Ray Observatory's public archives) of imaging various objects in space (up to $10^{23}$ cm). An analysis of the Redshift Surveys of galaxies and quasars suggests the possibility to draw the above similarity farther, up to $10^{26}$ cm. The phenomenon of skeletons is strengthened by the evidences which come from the resolution of the fine structure of luminosity of bright spots in the range $10^{-1}$-$10^{22}$ cm. The above analyses hint at the presence of a baryonic cold dark skeleton (BCDS) of the Universe, which -- in the entire range $10^{-5}$-$10^{26}$ cm -- may contain *ordinary* matter in a fractal condensed form like that in the above-mentioned dust skeletons and hail particles. The probable compatibility of the BCDS with the major cosmological facts (Hubble's expansion and cosmic microwave background) is suggested. Our former hypotheses (and the respective proof-of-concept studies) for the probable microscopic mechanisms of skeleton's assembling, chemical composition, and survivability in ambient hot plasmas are discussed briefly. The respective major cosmological implication is that the purely gravitational description of the large-scale structure of the Universe is likely to be appended with a contribution of quantum electromagnetism, presumably in the form of a skeleton self-assembled from tubular nanostructures (carbon nanotubes or similar nanostructures of other chemical elements).

e-mail: kuka@nfi.kiae.ru, rank@nfi.kiae.ru

## 1. Introduction. Extrapolating the phenomenon of skeletal structuring from laboratory electric discharges to space.

Observations of transverse-to-electric current, few-centimeters long straight filaments [1,2] of an anomalously long lifetime [3,4] in the plasma of gaseous electric discharge, a Z-pinch, lead to a hypothesis [3-5] that such filaments possess a microsolid skeleton which might be assembled during electric breakdown from wildly produced carbon nanotubes (or similar nanostructures of other chemical elements). Subsequent analysis [6] of electron micrographs of various types of dust deposits in the tokamak T-10 showed (i) the presence of tubular structures in the range several nanometers to several micrometers, (ii) the predicted [3-5] trend of assembling bigger tubules from smaller ones, and (iii) topological identity of cartwheel-like structures (often located in the edge cross-section of a tubule) in the above dust and in few-centimeters sized structures found [7,8] in the plasma images in small tokamaks. Further, the skeletons (tubules and cartwheels of millimeter-centimeter size) were found [8] at initial stage of discharge (e.g., before appearance of the discharge electric current) in tokamak, plasma focus, and vacuum spark. Here we show that the phenomenon of such skeletons may likely extend up to maximal lengths at which the Universe is a fractal.

This conjecture may be considered as an extension/modification of the hypothesis [9] of long-living filaments of electric current («plasma cables») which form electric circuits in space. The hypothesis [9] was supported, at a rough qualitative level [1,10], by the similarity of networking of filaments (of luminosity) in laboratory plasmas and space. At this point, the resolution of the fine structure of filaments and their networks in space, Earth's atmosphere and laboratory electric discharges enables us to indicate phenomena (namely a cartwheel-like network of filaments, an «electric torch-like» structure of bright spots, etc.) which suggest that in a very broad range of length scales (namely from nanoscale fractal dust [6] to large galaxies ~$10^5$ light-years across and, probably, even superclusters of galaxies ~$10^8$ light-years across) the filaments form the skeletons of the same distinctive topology that hints at the presence of a baryonic cold dark skeleton of the Universe.

## 2. Observations of similarity of structuring in a very broad range of length scales

### 2.1. Cartwheel-like structures in the range $10^{-5}$-$10^{23}$ cm.

We try to draw a bridge between laboratory experiments and space with presenting a short gallery of cartwheel-like structures -- probably the most inconvenient objects for being described universally for the *entire* range of length scales under consideration. In laboratory electric discharges [7,8] and respective dust deposits [6], the cartwheels are located either in the edge cross-section of a tubule or on an «axle-tree» filament, or as a separate block (the smallest cartwheels are of diameter less than 100 nm, see Figs. 2 and 3 in Ref. 6). Similar structuring of dramatically different size is found in the following typical examples of (i) big icy particles of a hail (Fig. 1A), (ii) a fragment of tornado (Fig. 1B), (iii) supernova remnant (Fig. 1C). One may obviously add to the last item of this list the Cartwheel galaxy, 150,000 light-years across and 500 million light-years away in the constellation Sculptor (.../gif/cartknot.gif [14]); the wheel-like supernova remnant G11.2-0.3 which is 40 light-years across and 25,000 light-years away in the constellation Sagittarius (.../cycle1/1227/1227_xray.tif [13]); a two-ring coaxial structure, with the inner ring of one light-year in diameter, in the centre of the Crab nebula which is 6,000 light-years away in the constellation Taurus (.../0052/0052_x-ray_lg.jpg [13], see also .../0052/crab-smooth-purple.jpg [13] for an adaptively smoothed image). Note that the cosmic wheel's skeleton (Fig. 1C) tends to repeat the structure of the icy cartwheel (Fig. 1A) up to details of its constituent blocks. In particular, some of radially directed spokes are ended with a tubular structure seen on the outer edge of the cartwheel. Moreover, in the edge cross-section of this tubular structure, the global cartwheel of the icy particle contains a smaller cartwheel whose axle is directed radially (see left lower window in Fig. 1A). Thus, there is a trend toward self-similarity (the evidences for such a trend in tubular skeletons found in the dust deposits are given in Ref. [6]).

Note that the images of Figures 1 and 2 are processed with the method of multilevel dynamical contrasting (MDC) [1,10]. As a rule, the structuring revealed with the help of MDC may then be easily recognised in the original, non-processed images (especially, for properly magnified high-resolution images).

## 2.2. Electric torch-like structures and self-illumination of skeleton in the range $10^{-1}$-$10^{22}$ cm.

Besides the distinctive topology (e.g., the cartwheels) of general layout of bright spots within skeletal structures, another evidence for the phenomenon of skeletons comes from the resolution of fine structure of luminosity around, at first glance, solitary bright spots. Here, the best evidence is the shining edge of a truncated straight filament which belongs to a skeletal network (see Figure 2). It seems that the straight blocks of skeletons may work as a guiding system for (and/or a conductor of) electromagnetic (EM) signals. Therefore, the open end of a dendritic electric circuit or a local disruption of such a circuit (e.g., its sparkling, fractures, etc.) may self-illuminate it to make it observable. This suggests a new interpretation to some mysterious events which involve binary systems (often quasi-symmetric ones) like, e.g., «colliding galaxies» and some dual nebulae. The first class may be represented by an «intergalactic «pipeline» (the dark, 20,000 light-years long string) of material flowing between two battered galaxies» 300 million light-years from Earth in the constellation Taurus (.../PR/2001/02/ [14]). The second class includes some butterfly-shaped nebulae (see .../nebulae.html [14]). The most relevant example here is the "Southern Crab Nebula" (He2-104), which looks like a sparkling in the fracture of a tubular structure (.../PR/1999/32/ [14]); less convincing examples are the "Papillon" Nebula with the bright spot in the centre of one of dual parts (.../pr/1999/23/ [14]) and the nebula NGC 2346 without any bright spot (.../PR/97/07/ [14]).

## 2.3. The signs of skeletal structuring at cosmological lengths, up to $10^{26}$ cm.

With increasing length scales, the self-illumination of the skeletal network in its certain, critical points continues working but the respective dramatic decrease of the average density of *hot, radiating* baryonic matter leads to observability of exclusively dim dotted imprints of skeletons, like e.g. mysterious dotted images of arcs and circles/ellipses. It appears that at largest observable lengths the more or less definite examples of distinctive topology, similar to that of Fig. 1, may be found only in the redshift surveys of *thin* slices of space (the redshift surveys are believed to provide a three dimensional distribution of galaxies, which may give, in particular, the side-on view on a thin conical slice of space). Figure 3 gives three examples of such structuring to continue the gallery of Figure 1. The original data are taken from three different projects, the two-degree-field Galaxy Redshift Survey (2dFGRS) [17], the Las Campanas Redshift Survey [18], and the 2dF Quasar Redshift Survey (2QZ) [19], which plotted in the redshift space the distribution of, respectively, some 140,000 and 20,000 galaxies (for redshifts $Z < 0.3$, i.e. at distances L up to 2.5 billion light-years away), and some 20,000 quasars ($Z < 3$, i.e. L ~ $1.5 \cdot 10^{10}$ light-years). The interpolation of visually correlated sequences of spots (e.g., by means of thickening the spots and subsequent smoothing the image) often gives various skeletal structures (namely, arcs, rings, straight filaments, sometimes the fragments of tubules and cartwheels) of various size and declination with respect to the observer. From optimistic viewpoint, a substantial part of the images may be reduced to a superposition of skeletal structures. (Note that such a structuring may be compatible with the gradual rise of the fractal correlation dimension D (see Table 1 in Ref. [20]) toward the limit of a continuous, non-fractal medium (D=3), with increasing the scale of observation, if the respective average density of skeletal structures tends to a certain limit.) Despite the structuring seen in Fig. 3 is obviously less reliable than that in Fig. 1, the correlation revealed makes it reasonable to suggest an extrapolation of our hypothesis farther to cosmological scales.

## 3. Cosmological implications of the above similarity: hypothesis of a baryonic cold dark skeleton (BCDS)

The immediate and most important implication of bringing all the above skeletons under one roof is that they may contain *ordinary* matter -- in a fractal condensed form, like that in the above-mentioned dust skeletons [6] – which, as suggested below, may work as a ***baryonic cold dark skeleton*** (BCDS).

### 3.1. Diminishing the controversy between luminosity and purely gravitational dynamics (the «dark matter» problem).

The certain rigidity of skeletal blocks, their enhanced mutual «viscosity» of a non-hydrodynamic, non-gravitational origin, and the presence of bright spots within skeletons, may produce a «skeletal» component of luminous objects. The visible objects which belong to this component may themselves move faster than predicted by their masses estimated from their luminosity and also may involve the gaseous/plasma component in a faster motion. These effects may diminish (or even avoid) the well-known controversy between apparent masses and their gravitational dynamics. The proposed reinterpretation is applicable to both historic sources of introducing a dark matter, namely the periphery of rotating galaxies and especially the clusters of galaxies because just at larger length scales the effect of skeletal component may be stronger.

### 3.2. Simultaneous darkness and opacity (a «dark filament» problem).

The *simultaneous* darkness and opacity of skeletal structures at cosmological lengths is expected from respective extrapolation of such a phenomenon at galactic and smaller length where this phenomenon is identifiable at the background of bright enough objects. These dark and opaque objects seem to be not an ordinary dust cloud because the presence of skeletal structures of such clouds is suggested by the presence of resolvable electric-torch structures like those in Fig. 2C. (We have to note that the phenomenon of «dark filaments», which was assumed [1,10] to be treatable in terms of the standard plasma physics, now may be related to BCDS.) Therefore, besides the rigidity, both the mass of such «dark» filaments and the luminous mass hidden by these opaque filaments may influence the dynamics of visible objects.

### 3.3. Probable compatibility of BCDS with two major cosmological facts.

Assuming a skeletal, rigid body structuring at *cosmological* lengths allows this hypothetical skeleton (excluding its luminous, negligibly small fraction) to be closer to thermodynamic equilibrium and be colder than any hypothetical dark baryonic matter in other phase states (liquid and/or gaseous) because in the skeleton the role of individual degrees of freedom (including the temperature) is minimal. This, in turn, suggests two implications. First, the cosmological skeleton which is actually not dark (as it should be for the baryonic matter) may be visible preferably as a thermally equilibrium (i.e. Planckian and isotropic) part in the entire observed radiation spectrum. Thus we immediately come to a hypothesis of thermal equilibrium between the skeleton and the cosmic microwave background radiation (CMBR) and to respective coldness of the overwhelming part of the skeleton. And second, the Hubble's expansion of the Universe may appear to be the expansion phase of a long-wavelength slow oscillation of a rigid body structure (a long-wavelength (spherical?) phonon). Correspondingly, the elasticity of the BCDS may be responsible for a global temporal antigravity which, in the purely gravitational description, would correspond to introduction of the cosmological constant.

## 4. Hypotheses for the probable mechanisms of observed structuring

The above picture of BCDS encounters significant obvious obstacles like the survivability of skeleton in hot luminous points, the compatibility of the certain rigidity of the BCDS not only with the Hubble's expansion of the Universe but also with the probable initial, «inflationary» phase of BCDS (i.e. with the very ability of self-assembling of BCDS, e.g., under condition of the Big Bang-like initial conditions). According to the assumed universality of the above phenomenon of skeletons, we have to treat the above problems in the entire range of length scales under consideration and to start from molecular length scale as it follows from the presence of *fully rigid-body* skeletons of certain topology in the dust deposits in laboratory discharges [6,21] and hail [11] (Fig. 1A).

### 4.1. Probable microscopic picture of skeleton's assembling and chemical composition.

The predictability [3-5] of discovering the microdust skeletons [6,21] was based on appealing [4,5] to the exceptional electrodynamic properties of their hypothetical building blocks -- first of all, the ability of these blocks to facilitate the electric breakdown in laboratory discharges and to assemble the micro- and macroskeletons. The carbon nanotube [22] (CNT) or similar nanostructures of other chemical elements have been suggested [3-5] to be such blocks. The self-assembling of skeletons was suggested to be based dominantly on *magnetic* phenomena and assumes the following simultaneous processes, namely ($\alpha$) externally driven expansion/inflation of the self-assembling networks (in laboratory discharges, there is an inflow of magnetic field from the external electric circuit, which is especially intense at the initial stage of discharge), ($\beta$) sticking of the blocks together -- due to trapping of magnetic flux by CNTs and/or their assemblies and respective mutual magnetic dipole attraction; and due to experimentally verified trend [23] in magnetically confined plasmas to form the so-called force-free magnetic configuration which sustains a balance between longitudinal and transverse confinement/self-attraction in the plasma column as well as in the electric current filament, and ($\gamma$) *partial* solidification due to welding of blocks by the passing electric current. We believe the mechanisms ($\alpha$),($\beta$), and ($\gamma$) to select ultimately the structures of a matter-saving, survivable geometry – first of all, tubules, cartwheels, and their combinations.

The indications on plausibility of the anomalous magnetism and, in particular, on the ability of CNTs and/or their assemblies to trap and almost dissipationlessly hold a magnetic flux, with the specific magnetization high enough to stick the CNTs together, come from observations of superconductor-like diamagnetism in self-assemblies of CNTs (inside *non-processed* fragments of cathode deposits) at room temperatures [24] and in artificial assemblies at 400 K [25]. (See also observations [26] of unexpected ferromagnetism of a *pure* carbon, rhombohedral $C_{60}$, with a Curie temperature near 500 K).

Further, the plausibility of simultaneous opacity and darkness of «dark filaments» is supported by the existence of ultra-disperse materials of anomalous blackness. Interestingly, the submicron agglomerates of the Carbon Black particles may have the structure similar to that of the submicron agglomerates found [21] in the carbonaceous dust deposits in tokamak T-10 (e.g., one may recognize the cartwheel-like structuring in the electron micrographs [27] of the carbon black).

**4.2. On the probable compatibility of skeletal structuring with explosive dynamics (a planar filamentary explosion as a mechanism of cartwheel's birth, a parachute-like expansion of a dendritic network as a mechanism of BCDS's birth).**

The indications on the compatibility of rigidity of certain blocks with the expansion (or inflationary dynamics) both of the entire skeleton and its certain parts come partly from the presence [28] of typical skeletal structures in the x-ray images of one of the most bursty phenomenon in the laboratory, namely plasma corona produced by the irradiation of a solid target with a short laser pulse, and partly from available rare data on tracing the dynamics of cosmic explosions. In the latter case of a young star system, XZ Tauri, some 0.01 light-years across and 450 light-years from Earth in the Taurus-Auriga molecular cloud (.../PR/2000/32/ [14]), we found the signs of a two-dimensional expansion (it's worth to call this a «planar explosion») which seems to be an inflationary production of the cartwheel-like structures on the common axle and is compatible with the interplay of the above mechanisms ($\alpha$) and ($\beta$). One more example of the distinct planar structuring is a «large thin equatorial disk» (.../PRC96-23a [14]) less than one light-year across around the star Eta Carinae (.../EtaCarC.jpg [14]), 8,000 light-years away, which has been released as a supernova explosion.

Regardless of specific mechanism of inflation/explosion (either Coulombic, as resulted from an electrostatic instability, or nuclear energy release, from gravitational instability and collapse, or any other), the ($\alpha,\beta,\gamma$) mechanism may give a parachute-like expansion of a dendritic network (namely, a parachute with the «liquid» strops and the localised explosive sources of dendricity). Note that the typical block of skeletons, the cartwheel on an axle and the tubule with the central rod and the cartwheel in the edge cross-section, are both the dendrites (for the respective examples in the fractal dust deposits -- the skeletons composed of tubular nanofibers -- see Refs. [6,21]). The saturation of the expansion may manifest itself implicitly, e.g., like local termination of the dendritic growth because of local exhaust of the proper material (this may give the open, «shining» ends of the network, see Fig. 2). The examples of the products of a filamentary explosion seem to be the nebulae which, under high resolution imaging, appear to possess a distinct skeleton (see, e.g., the «Eskimo» Nebula, .../pr/2000/07/ [14]) but look like a constellation if the brightest spots are retained in the image (via contrasting the image). To some extent, such a closed structure may be considered as a very rough model of a closed skeletal Universe in a miniature.

**4.3. On the survivability of skeletons in hot plasmas.**

The solution to the problem of survivability of skeleton in hot plasmas has been suggested within the framework of the problem of non-local (non-diffusive) transport of heat observed in high-temperature plasmas for controlled thermonuclear fusion. The microsolid skeletons were suggested [7] to be self-protected from an ambient high-temperature plasma by thin vacuum channels self-consistently sustained around the skeletons by the pressure of high-frequency (HF) electromagnetic waves, thanks to the skeleton-induced conversion of a small part of the incoming slow, quasi-static magnetic field (poloidal, in tokamaks, or azimuthal, in Z-pinches) into HF waves of the TEM type (a «wild cable» model, see P2_028 in Ref. [7]). This model allows to evaluate the width and length of vacuum channels around straight blocks of skeletons from the measurements of HF electric fields, both inside and outside the plasma column. The respective results [7,15] for the case of tokamak T-10 and gaseous Z-pinch reasonably agree with visible dimensions of observed straight blocks of skeletons.

## 5. Conclusion

The major cosmological implication of the above considerations and hypotheses is that the purely gravitational description of the large-scale structure of the Universe is likely to be appended with a contribution of the quantum electromagnetism, presumably in the form of a skeleton self-assembled from carbon nanotubes or similar nanostructures. Note also that in the above hypotheses, the integrity and continuity of BCDS is based on the *quanticity* (specifically, quantum electromagnetism) of *long-range* bonds, unlike the (i) commonly recognised quanticity of exclusively *short-range* electromagnetic processes (namely, elementary radiative-collisional processes, like e.g. radiation emission in spectral lines of atoms excited by particle's impact) and (ii) *classicality* of long-range bonds in the hypothetical astrophysical «plasma cables» [9]. Of course, in a broader frame, one should allow the integrity of skeletons (and the respective long-range quanticity) to come from physics beyond the frame of quantum electromagnetism, i.e. residing at a deeper level (i.e. smaller lengths) and ultimately manifested in the form of BCDS suggested above.

*The present work has been carried out by the authors in their spare time.*

**Acknowledgements.** We thank V.I. Kogan for invariable support and encouragement, all our colleagues who valuably contributed to former stages of our research (see survey papers [4,8,28]) by presenting the original databases and collaborating with us, especially to B.N. Kolbasov and P.V. Romanov. The partial financial support from the Russian Federation Ministry for Atomic Energy and the Russian Foundation for Basic Research is acknowledged.

# REFERENCES

1. Kukushkin, A. B. & Rantsev-Kartinov, V. A. Dense Z-pinch plasma as a dynamical percolating network: from laboratory plasmas to a magnetoplasma universe. *Laser and Part. Beams* **16,** 445-471 (1998).
2. Kukushkin, A. B. & Rantsev-Kartinov, V. A. Observations of a dynamical percolating network in dense Z-pinch plasmas. *Rev. Sci. Instrum.* **70,** 1421-1426 (1999).
3. Kukushkin, A. B. & Rantsev-Kartinov, V. A. Microsolid tubular skeleton of long-living filaments of electric current in laboratory and space plasmas. in *Proc. 26-th Eur. Phys. Soc. conf. on Plasma Phys. and Contr. Fusion* (Maastricht, Netherlands, June 1999) 873-876 (http://epsppd.epfl.ch/Maas/web/pdf/p2087.pdf).
4. Kukushkin, A. B. & Rantsev-Kartinov, V. A. Long-living filamentation and networking of electric current in laboratory and cosmic plasmas: from microscopic mechanism to self-similarity of structuring. in *Current Trends in International Fusion Research, Proc. 3$^{rd}$ Symposium,* Washington, D.C., 1999 (ed. Panarella, E.) (NRC Research Press, Ottawa, Canada, 2002) pp.107-135 (to be published).
5. Kukushkin, A. B. & Rantsev-Kartinov, V. A. Filamentation and networking of electric currents in dense Z-pinch plasmas. in *Fusion Energy 1998 (Proc. 17-th IAEA Fusion Energy Conference, Yokohama, Japan, 1998*) **3,** 1131-1134 (IAEA, Vienna, 1999, IAEA-CSP-1/P), (http://www.iaea.org/programmes/ripc/physics/pdf/ifp_17.pdf).
6. Kolbasov, B. N., Kukushkin, A. B., Rantsev-Kartinov, V. A. & Romanov, P. V. Similarity of micro- and macrotubules in tokamak dust and plasma. *Phys. Lett. A,* **269**, 363-367 (2000).
7. Kukushkin, A. B. & Rantsev-Kartinov, V. A. Wild cables in tokamak plasmas. in *Proc. 27-th Eur. Phys. Soc. conf. on Plasma Phys. and Contr. Fusion*, Budapest, Hungary, June 2000 (http:// epsppd.epfl.ch/Buda/pdf/p2_029.pdf; .../p2_028.pdf).
8. Kukushkin, A. B. & Rantsev-Kartinov, V. A. Long-lived filaments in fusion plasmas: review of observations and status of hypothesis of microdust-assembled skeletons. in *Current Trends in International Fusion Research, Proc. 4$^{th}$ Symposium,* Washington D.C., 2001 (eds. C.D. Orth, E. Panarella, and R.F. Post) (NRC Research Press, Ottawa, Canada, 2002) (to be published); (see also preprint http://xxx.lanl.gov/pdf/physics/0112091).
9. Alfvén, H. *Cosmic Plasma* (D. Riedel Publ. Co., Dordrecht, Holland, 1981).
10. Kukushkin, A. B. & Rantsev-Kartinov, V. A. Self-similarity of plasma networking in a broad range of length scales: from laboratory to cosmic plasmas. *Rev. Sci. Instrum.* **70,** 1387-1391 (1999).
11. Australian Severe Weather, http://australiasevereweather.com/photography/photos/1996/0205mb11.jpg; .../0205mb12.jpg.
12. National Oceanic & Atmospheric Administration Photo Library, Historic National Weather Service Collection, http://www.photolib.noaa.gov/historic/nws/images/big/wea00216.jpg.
13. Chandra X-Ray Observatory - Home. http://chandra.harvard.edu/photo/...
14. Space Telescope Science Institute - Hubble Space Telescope Public Information, http://oposite.stsci.edu/pubinfo/...
15. Kukushkin, A. B. & Rantsev-Kartinov, V. A. Wild cables in a Z-pinch and plasma focus. in *Proc. 27-th Eur. Phys. Soc. conf. on Plasma Phys. and Contr. Fusion*, Budapest, Hungary, June 2000 (http://epsppd.epfl.ch/Buda/pdf/p2_051.pdf).
16. Vonnegut, B., & Weyer, J. Luminous phenomena in nocturnal tornadoes. *Science* **153**, 1213 (1966) (see also a short survey of witnesses to luminous tornadoes in a paper by E. Lewis at http://www.padrak.com/ine/ELEWIS3.html).
17. The 2dF Galaxy Redshift Survey, http://www.mso.anu.edu.au/2dFGRS/...

18. Shectman, S. A. *et. al.* The Las Campanas redshift survey. *Astrophys. J.* **470**, 172-188 (1996).
19. The 2dF QSO Redshift Survey, http://www.2dfquasar.org/wedgeplot.html.
20. Wu, K. K. S., Lahav, O., & Rees, M. J. The large-scale smoothness of the Universe. *Nature* **397**, 225-230 (1999).
21. Kolbasov, B. N., Kukushkin, A. B., Rantsev-Kartinov, V. A. & Romanov, P. V. Skeletal dendritic structure of dust microparticles and of their agglomerates in tokamak T-10. *Phys. Lett. A,* **291**, 447-452 (2001).
22. Iijima, S. Helical microtubules of graphitic carbon. *Nature* **354,** 56-58 (1991).
23. Taylor, J.B., Relaxation and magnetic reconnection in plasmas. *Rev. Mod. Phys.* **58**, 741-763 (1986).
24. Tsebro, V.I. & Omel'yanovskii, O.E. Undamped currents and magnetic field trapping in a multi-connected, carbon nanotube structure. *Phys. Usp.* **43,** 847 (2000).
25. Zhao, G. & Wang, Y.S. Possible superconductivity above 400 K in carbon-based multiwall nanotubes. Preprint cond-mat/0111268 (at xxx.lanl.gov) (to be published in Phil. Mag. B).
26. Makarova, T.L., *et. al.* Magnetic carbon. *Nature* **413**, 716-718 (2001).
27. Dramstad, H., Roy, C. & Kaliaguine, S. Characteristics of carbon black... in *Proc. 23$^{rd}$ Biennal Conference on Carbon*. Pennsylvania State U., USA, 1997 (see http://www.gch.ulaval.ca/~darmstad/sem.html, Fig. 2).
28. Kukushkin, A. B. & Rantsev-Kartinov, V. A. Skeletal structures in high-current electric discharges and laser-produced plasmas: observations and hypotheses. in *Advances in Plasma Phys. Research* (Ed. F. Gerard, NovaPublishers, N.Y., 2002), (see http://www.novapublishers.com/detailed_search.asp?id=1-59033-200-8).

**Similarity of structuring in the range $10^{-5}$ cm to $10^{23}$ cm hints at a baryonic cold dark skeleton of the Universe,**
A.B. Kukushkin & V. A. Rantsev-Kartinov

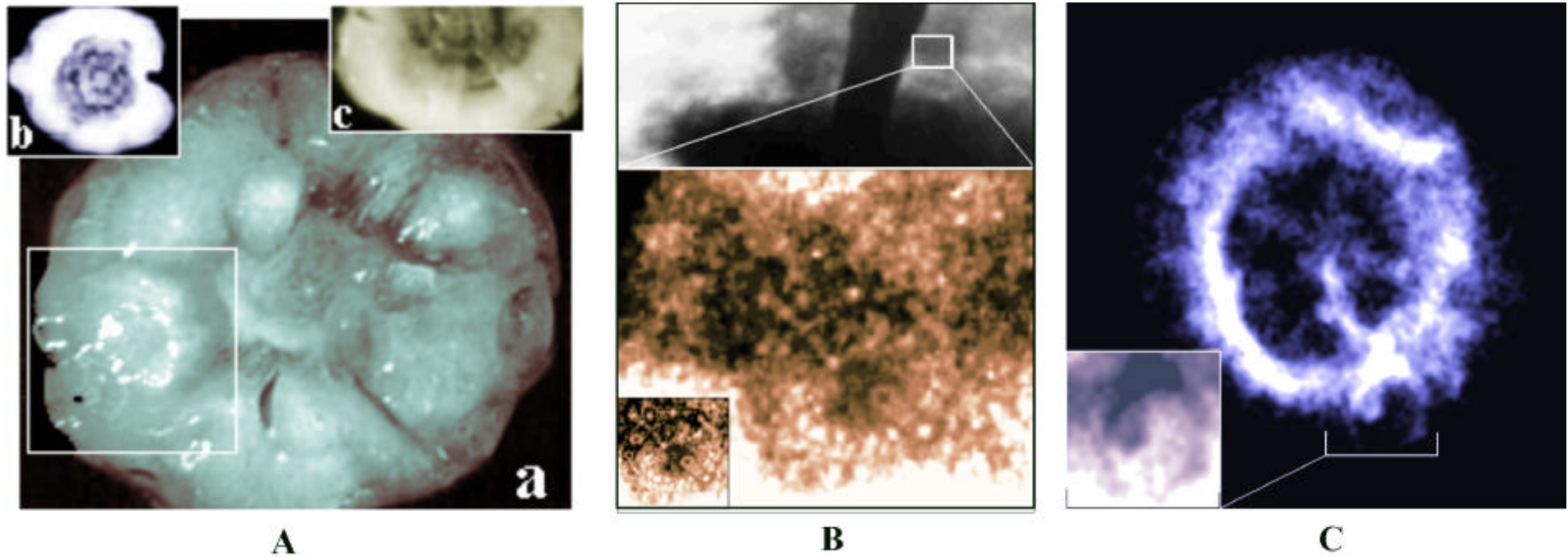

Figure 1. The cartwheel-like structures at different length scales. **A**, Big icy particles of a hail of diameter 4.5 cm (a), 5 cm (b), and 5 cm (c). The original images are taken from Ref. [11]. The frame in the left lower part of the image (a) is contrasted separately to show an elliptic image of the edge of the radially directed tubular structure. The entire structure seems to contain a number of similar radial blocks. A distinct coaxial structure of the cartwheel type is seen in the central part of image (b). Image (c) shows strong radial bonds between the central point and the «wheel». **B**, Top section: A fragment of the photographic image [12] of a massive tornado of estimated size of some hundred meters in diameter. Bottom section: A fragment of the top image shows the cartwheel whose slightly elliptic image is seen in the centre. The cartwheel seems to be located on a long axle-tree directed down to the right and ended with a bright spot on the axle's edge (see its additionally contrasted image in the left corner insert on the bottom image). **C**, «A flaming cosmic wheel» of the supernova remnant E0102-72, with «puzzling spoke-like structures in its interior», which is stretched across forty light-years in Small Magellanic Cloud, 190,000 light-years from Earth (.../snrg/e0102electricbluet.tif [14]). The radially directed spokes are ended with tubular structures seen on the outer edge of the cartwheel. The inverted (and additionally contrasted) image of the edge of such a tubule (marked with the square bracket) is given in the left corner insert (note that the tubule's edge itself seems to possess a tubular block, of smaller diameter, seen on the bottom of the insert). Thus, the cosmic wheel's skeleton tends to repeat the structure of the icy cartwheel up to details of its constituent blocks.

**Similarity of structuring in the range $10^{-5}$ cm to $10^{23}$ cm hints at a baryonic cold dark skeleton of the Universe,**
A.B. Kukushkin & V. A. Rantsev-Kartinov

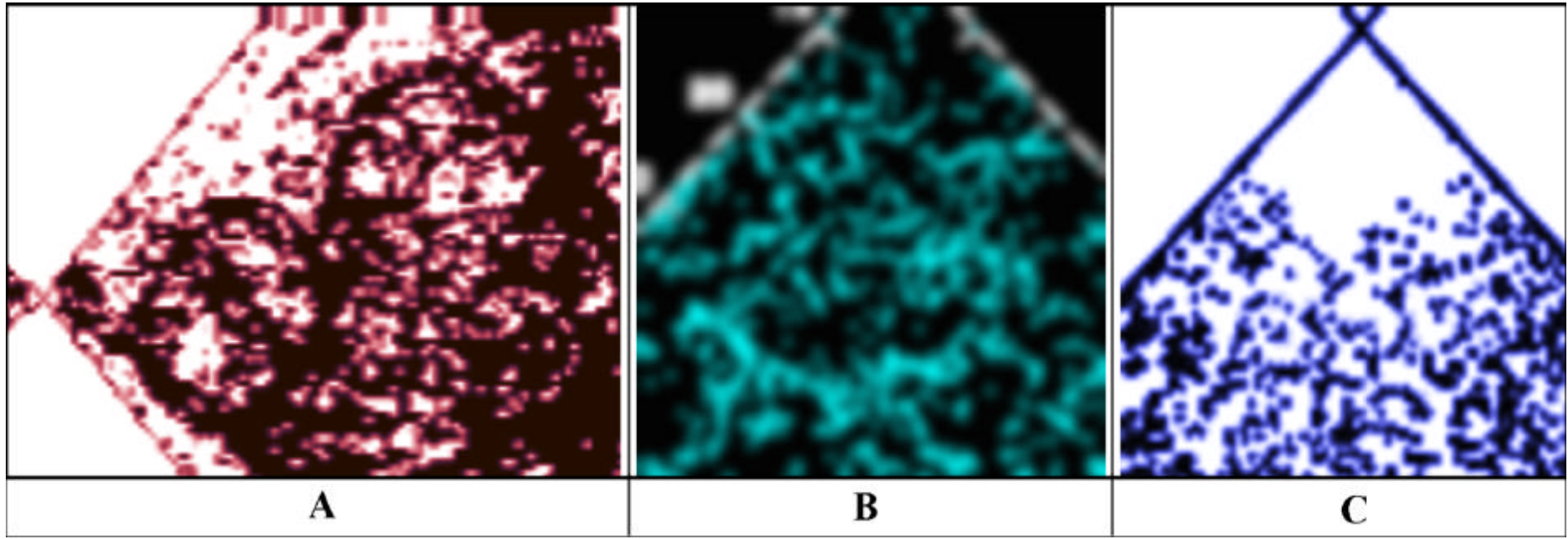

Figure 3. The fragments of distribution of galaxies/quasars in the redshift space, which is believed to give a side-on view of a conical slice of space (the Earth is located on the cone's vertex). Original data are taken, respectively, from the 2dF Galaxy Redshift Survey [17], the Las Campanas Redshift Survey [18], and the 2dF QSO Redshift Survey [19]. **A**, A fragment of the projected distribution [17] of the galaxies in the South Galactic Pole strip (4° thick slice is centred at declination -27.5°), as a function of redshift Z and right ascension. The lower border of the cone reaches the bottom of the figure at $Z \sim 0.027$ (or, equivalently, at a distance $\sim 2.7 \cdot 10^8$ light years). The slight increase of spots' size in original image at .../Public/Pics/2dFzcone.gif [17] gives, e.g., the appearance of elliptic image of a circular (or at least, an arc-like) structure; **B**, A fragment of similar distribution of the galaxies [18] (1.5° thick slice is centred at declination -45° in the South Galactic Pole strip, see red points in the coloured image at http://www.astro.ucla.edu/~wright/lcrs.html)). The left border of the cone crosses the left hand side of the figure at a distance $\sim 1.5 \cdot 10^9$ light years. Thickening of the spots and subsequent smoothing of the image gives a circle and straight radial filaments. **C**, A fragment of similar distribution of quasars [19] (2-5° thick slice is centred at declination -30° in the South Galactic Pole strip) contains the circles with central point and the arc-like structures. The right border of the cone crosses the right hand side of the figure at a distance $\sim 5 \cdot 10^9$ light years.